\documentclass[11pt,onecolumn,amssymb,nofootinbib]{revtex4}
\usepackage{amsmath, amsthm, amscd, amssymb}
\usepackage{bm}
\usepackage{bbm}
\usepackage{slashed}

\begin{document}

\title{\bf Analysis of a gauged model with a spin-$\frac{1}{2}$ field directly coupled to a Rarita-Schwinger spin-$\frac{3}{2}$ field}

\author{Stephen L. Adler}
\email{adler@ias.edu} \affiliation{Institute for Advanced Study,
Einstein Drive, Princeton, NJ 08540, USA.}

\begin{abstract}
We give a detailed analysis of  an abelianized  gauge field  model in which a Rarita-Schwinger spin-$\frac{3}{2}$ field is directly coupled to a spin-$\frac{1}{2}$ field.
The model permits a perturbative expansion in powers of the gauge field coupling, and from the Feynman rules for the model we calculate the chiral anomaly.
\end{abstract}

\maketitle

\section{Introduction}

A long-standing question in the theory of Rarita-Schwinger spin-$\frac{3}{2}$ fields is whether, apart from the context of $O(N)$ supergravity theories \cite{supergrav}, they can be consistently gauged.  There are several motivations for investigating this question.  First, despite three decades of effort at grand unification of the
particle forces, a satisfactory theory has not been found.  The conventional assumption is that in formulating grand unification models, gauge anomalies must be canceled
within the fermionic spin-$\frac{1}{2}$ sector.  If spin-$\frac{3}{2}$ theories can be gauged, then additional avenues for model building are opened, in which anomalies
cancel between the spin-$\frac{3}{2}$  and spin-$\frac{1}{2}$ sectors.  Second, the literature on gauge anomalies in spin-$\frac{3}{2}$ theories is perplexing. While there
are a number of calculations of gauge anomalies in the spin-$\frac{3}{2}$  case \cite{spin3/2anom} for general gauge groups including $SU(N)$, there are no supergravity theories in which $SU(N)$ is gauged, raising the question of whether there is some version of the spin-$\frac{3}{2}$ theory that admits a consistent
gauging with a general gauge group.  Finally, the author has proposed \cite{adlersu8} a  unification model based on an $SU(8)$ gauge group, in which gauge anomalies
are canceled between the fermionic  spin-$\frac{3}{2}$  and spin-$\frac{1}{2}$ sectors, raising in this specific context the question of whether spin-$\frac{3}{2}$
theories can be consistently gauged.

The classic papers of Johnson and Sudarshan \cite{johnson} and Velo and Zwanziger \cite{velo} on spin-$\frac{3}{2}$ theory found  problems,  including
non-positivity of anti-commutators and superluminal modes, in the massive field case.  Since from a modern point of view masses are never put in ``by hand'', but
instead must arise from spontaneous symmetry breaking, the author undertook a detailed study \cite{adlerRS1}, \cite{adlerRS2} of massless spin-$\frac{3}{2}$ theory, making extensive use of the left chiral reduction to facilitate the calculations.  This work showed that in the massless case all modes are luminal, and that by adding an auxiliary
field, the gauged theory could be extended to have a full off-shell fermionic gauge invariance.  A detailed investigation of the extended theory by the author, Henneaux, and Pais \cite{henneaux} showed that the extended model still has problems.  When covariant gauge fixing is correctly implemented, the extended model has non-positive anti-commutators, and more seriously from the point of view of quantization, the Dirac brackets in the constrained theory have a weak field singularity in the
massless case, that is shifted from the spin-$\frac{3}{2}$ bracket to the auxiliary field bracket, but is not removed.

With this motivation, in the present paper we study a gauged model based on the specific manner in which Rarita-Schwinger fields enter into the model of \cite{adlersu8}, involving
a direct coupling of the spin-$\frac{3}{2}$ field to a spin-$\frac{1}{2}$ field, which preserves chiral symmetry and so is not a conventional mass term.  We find that
although the problems of non-positive anti-commutators and superluminal modes appear in this model, as discussed in more detail in the concluding section, the more serious problem of a weak field singularity is eliminated.  Thus the quantized model  admits a perturbation expansion in the gauge field coupling, and allows calculation of the gauge anomaly.

This paper is organized as follows.  The first part, comprising Secs. 2-9, introduces the coupled model and reprises in this context the analysis of Refs. \cite{adlerRS1}, \cite{adlerRS2},  and \cite{henneaux}.  In Sec. 2 we recall how the chiral symmetry preserving spin-$\frac{3}{2}$ to spin-$\frac{1}{2}$ coupling, but not a chiral symmetry
breaking mass term, arises in the model of \cite{adlersu8}.  In Sec. 3 we abstract from this model a simplified, abelianized version, and discuss the equations of motion
and constraints in four-component form.  In Sec. 4 we give the left chiral reduction of the model.  In Sec. 5 we analyze the canonical momenta, constraint brackets,
and counting of degrees of freedom.  Zero external field plane wave solutions are studied in Sec. 6, and the wave velocity in an external gauge field is studied by the
method of characteristics in Sec. 7.  In Sec. 8 we derive the canonical brackets, the Dirac brackets, and the Hamiltonian equations of  motion, and in  Sec. 9 we formulate
path integral quantization.

The second part of this paper, comprising Secs. 10-13,  is devoted  to the calculation of the chiral anomaly in the model.  In Sec. 10 we give a preliminary, heuristic discussion of the ghost contribution to the chiral anomaly, which suggests that it is equal to -1 times the standard spin-$\frac{1}{2}$ chiral anomaly.  In Sec. 11 we derive the Feynman rules for propagators and vertex parts in the coupled model, and give the Ward identities that these obey in Sec. 12.  Section
13 is devoted to chiral anomaly calculations.  In Sec. 13A, we derive the standard spin-$\frac{1}{2}$ chiral anomaly by the shift method. In Secs. 13B and 13C we give
two alternative calculations of the fermion loop contribution to the coupled model chiral anomaly.  These calculations show this  is equal to 5 times the  standard spin-$\frac{1}{2}$ chiral anomaly, just as in the uncoupled limit of the model.  Thus, if the ghost analysis of Sec. 10 is confirmed by further study, the
chiral anomaly in the coupled model takes a value of 4 times that of the standard spin-$\frac{1}{2}$ chiral anomaly, and evolves continuously from the the uncoupled limit
to the case of nonzero coupling. A brief discussion of the results of this paper, and directions for future work, are given in Sec. 14.  Appendix  A contains a summary
of notational conventions and useful identities, and Appendix B gives details of the calculation of plane wave solutions.

\section{The non-abelian model}
Although most of this paper focuses on an abelianized model, we begin by sketching the non-Abelian $SU(8)$ gauge field model \cite{adlersu8} from
which it is abstracted.  This model has the field content shown in Table 1. Because of the group representation content, no $SU(8)$ invariant Yukawa type coupling
of the Rarita-Schwinger field $\psi_{\mu}$ to the scalar field $\phi$ can be constructed.  That is, if one attempts to write down
a term trilinear in $\overline{\psi}_{\mu}$, $\psi_{\mu}$, and $\phi$, which would become a Rarita-Schwinger mass after the scalar field acquires a vacuum expectation
value, there is no way to the contract the internal symmetry indices shown in Table 1 to form an $SU(8)$ invariant.  So gauge symmetry forbids the appearance of a conventional Rarita-Schwinger mass term.  However, one
can \cite{adler3} write down a gauge invariant trilinear coupling of the Rarita-Schwinger field and the scalar field to a linear combination $\lambda$ of the two
representation $\overline{28}$ fields $\lambda_{1,2}$ of the model,
\begin{equation}\label{RS}
\overline{\lambda}^{[\alpha\beta]} \gamma^{\nu} \psi_{\nu}^{\gamma} \phi^*_{[\alpha \beta \gamma]}
\end{equation}
and its conjugate,
which becomes a proto-mass term
 \begin{equation}\label{RS1}
\overline{\lambda}^{[\alpha\beta]} \gamma^{\nu} \psi_{\nu}^{\gamma} \overline{\phi}^{\,*}_{[\alpha \beta \gamma]}
-\overline{\psi}_{\nu\gamma}\gamma^{\nu}\lambda_{[\alpha\beta]}\overline{\phi}^{[\alpha\beta\gamma]}~~~
\end{equation}
when the scalar field $\phi$ develops an expectation $\overline{\phi}$.  We use the term ``proto-mass''
because the coupling of Eq. \eqref{RS1}  still couples left-chiral field components to left-chiral components, rather than to
right-chiral components as in a Dirac mass term.  For $\psi_{\mu}$ to develop a mass that couples  left-chiral to
right-chiral components, a further stage of dynamical symmetry breaking (not the focus of this paper) will be
needed.  We note that both Eqs. \eqref{RS} and \eqref{RS1} already break the free Rarita-Schwinger field
fermionic gauge symmetry $\psi_{\mu} \to \psi_{\mu} + \partial_{\mu}\epsilon$, so there is no change in the
degree of freedom counting when the model is gauged by replacing the derivative $\partial_{\mu}$ by the
gauge-covariant derivative $D_{\mu}=\partial_{\mu} +g A_{\mu}$, with $A_{\mu}$ an anti-Hermitian gauge field.
This avoids the vexing issue of a discontinuity in the degree of freedom counting when a Rarita-Schwinger field
is gauged, which is analyzed in detail in \cite{henneaux}.

The $SU(8)$ gauge model with field content set out in Table 1 has a number of interesting properties,
which make it special, and possibly unique.  (1) Although not supersymmetric for nonzero gauge coupling, the model has boson-fermion balance, that is, the numbers of boson and fermion degrees of freedom are the same.  This can play a role in cancellation of leading order contributions to the
cosmological constant.   (2)  The model cancels $SU(8)$ gauge anomalies between the two multiplets separated by the horizontal line in Table 1.  The multiplet containing the graviton and gravitino is anomalous by itself; the second multiplet containing two $\overline{28}$ fermions and a complex spin 0 boson is needed to cancel anomalies.  (3)  The model is highly frustrated, in the sense that many couplings of the spin 0 field to the fermions allowed by dimensional counting are forbidden by the $SU(8)$ gauge symmetry. Thus, as explained above, a conventional Rarita-Schwinger mass term is forbidden, with only a chirality-preserving coupling of the spin-$\frac{3}{2}$ field to spin-$\frac{1}{2}$ fields allowed.
Additionally, no Yukawa couplings of the spin 0 field to the spin-$\frac{1}{2}$ fields are allowed by the
gauge symmetry.  This means that the model has only one dimensionless coupling constant, the coefficient of the coupling term of Eqs. (1) and (2).   Thus the theory is highly '`calculable'' in the sense that  most of its phenomenological consequences will be calculable numbers  as opposed to renormalizable couplings and masses.   We suspect also that frustration  may provide a way to generate a scale hierarchy in the absence of full supersymmetry, since many couplings in the
effective phenomenological Lagrangian can only appear in higher orders of perturbation theory.
(4)  The gauge group $SU(8)$ is just large enough to accommodate the standard model gauge group with an
additional ``technicolor'' gauging, which can lead to generation of a composite standard model Higgs field.
(5)  The model contains a hint of a possible connection with the $E_8$ root  lattice, since the helicity counts
in the two multiplets of Table 1 are 128 and 112, suggesting that there may be a relation to the 240 shortest vectors of the $E_8$ lattice.   This is a direction to be explored to find a further unification of the quantum field theoretic  $SU(8)$ model with general relativity.   Whether this possibility, and others stemming from the features just enumerated, is realized can only be determined by a detailed study of the consequences of the model, which is the program of which this paper is a part.

\begin{table} [ht]
\caption{Field content of the model of \cite{adlersu8}.  Square brackets indicate complete antisymmetrization of the enclosed indices.  The indices $\alpha,\beta,\gamma$ range from 1 to 8, and the index $A$ runs from 1 to 63. The top and bottom sections of the table each
contain bosons and fermions satisfying the requirement of boson-fermion balance, with the helicity counts for top and bottom
128 and 112 respectively.}
\centering
\begin{tabular}{ c c c c c}
\hline\hline
field~~~ & spin~~~ & $SU(8)$ rep.~~~ & helicities~~~  \\
\hline
$h_{\mu \nu}$ & 2 & 1 & 2\\
$\psi_{\mu}^{\alpha}$ & Weyl \,3/2 & 8 & 16  \\
$A_{\mu}^A$ & 1 & 63 & 126 \\
$\chi^{[\alpha\beta\gamma]}$&Weyl \, 1/2 & 56 & 112 \\
\hline
$\lambda_{1[\alpha \beta]}$&Weyl\, 1/2&$\overline{28}$ & 56 \\
$\lambda_{2[\alpha \beta]}$&Weyl \,1/2&$\overline{28}$ & 56 \\
$\phi^{[\alpha\beta\gamma]}$& complex\, 0& 56 & 112 \\
\hline\hline
\end{tabular}
\label{fieldcontent}
\end{table}

\section{The abelianized model}

We now abstract from the non-Abelian model discussed in Sec. 2 an abelianized model, in which the $SU(8)$ gauge field of Table 1 is replaced by
an Abelian gauge field $A_{\mu}$, which is used to gauge the Rarita-Schwinger field $\psi_{\mu}$ and the spin-$\frac{1}{2}$ field $\lambda$.
To reflect the properties of the non-Abelian model of Sec. 1, we include no bare mass term for the Rarita-Schwinger field, but incorporate a direct
coupling of  $\psi_{\mu}$ to $\lambda$ as in Eq. \eqref{RS1}, relabeling the scalar field expectation $\overline{\phi}$, which now carries no
internal symmetry indices, as a mass $m$.  Thus, using the fermion field normalization convention of the text of Freedman and Van Proeyen \cite{freedman}
(which differs by an overall factor of 2 in the Lagrangian from the convention we used in our earlier papers), we start from the Lagrangian density
\begin{align}\label{covariant}
S=&S(\psi_{\mu})+S(\lambda)+S_{\rm interaction}~~~,\cr
S(\psi_{\mu})=&\int d^4x \overline{\psi}_{\mu} R^{\mu}~~~,\cr
R^{\mu}=&i\epsilon^{\mu\eta\nu\rho}\gamma_5\gamma_{\eta}D_{\nu}\psi_{\rho}
=-\gamma^{\mu\nu\rho}D_{\nu}\psi_{\rho}~~~,\cr
D_{\nu}\psi_{\rho}=&(\partial_{\nu}+gA_{\nu})\psi_{\rho}~~~,\cr
\overline{\psi}_{\mu}=&\psi^{\dagger}_{\mu} i \gamma^0~~~\cr
S(\lambda)=&-\int d^4x \overline{\lambda} \gamma^{\nu}D_{\nu}\lambda~~~,\cr
D_{\nu}\lambda=&(\partial_{\nu}+gA_{\nu})\lambda~~~,\cr
\overline{\lambda}=&\lambda^{\dagger} i \gamma^0~~~,\cr
S_{\rm interaction}=&m\int d^4x(\overline{\lambda}\gamma^{\nu}\psi_{\nu}-
\overline{\psi}_{\nu}\gamma^{\nu}\lambda)~~~.
\end{align}
Varying the action with respect to $\overline{\psi}_{\mu}$ and $\overline{\lambda}$ we get
the Euler-Lagrange equations of motion
\begin{align}\label{eqmo}
R^{\mu}=&m\gamma^{\mu} \lambda~~~,\cr
\gamma^{\nu}D_{\nu}\lambda=&m\gamma^{\nu}\psi_{\nu}~~~.\cr
\end{align}
The 0 component of the first line of Eq. \eqref{eqmo} involves no time derivatives, and so  gives the primary constraint of the theory,
\begin{equation}\label{primary}
-i\gamma_5 \epsilon^{enr}\gamma_eD_n\psi_r=m\gamma^0\lambda~~~.
\end{equation}
We see that in the limit $m\to \infty$, this constraint forces $\lambda$ to vanish, a feature that will reappear later.
Since $(\gamma^0)^2=-1$, Eq. \eqref{primary}
can be rewritten so that the term containing $\lambda$ is a spinor rather than the direct
product of the time component of a four-vector with a spinor,
\begin{equation}\label{spinorform}
0= \tilde \chi\equiv -i \gamma_5 \gamma^0 \epsilon^{enr}\gamma_eD_n\psi_r-m\lambda~~~,
\end{equation}
a form that will be used in Sec. 10 in discussing how the constraints enter the path integral.\footnote{The $\psi_r$ term in Eq. \eqref{spinorform} is a linear combination
of a Lorentz spinor and a Lorentz tensor-spinor, but using the $\psi_{\mu}$ equation of motion in Eq. \eqref{eqmo} it is easy to see that the latter vanishes identically,
and so does not give an additional constraint:  Form $T^{\mu\nu}\equiv  \gamma^{\nu}R^{\mu} + \gamma^{\mu} R^{\nu}= 2 m \eta^{\mu\nu}\lambda$.  The Lorentz tensor-spinor
part of $T^{\mu\nu}$ is $T^{\mu\nu}-\frac{1}{4} \eta^{\mu\nu} T^{\theta}_{\theta} =0$.  But $T^{00}$ equals $-2$  times the  $\psi_r$ term in Eq. \eqref{spinorform}, so
the Lorentz tensor-spinor part of Eq. \eqref{spinorform}  vanishes by virtue of the Euler-Lagrange equations.}

Contracting $D_{\nu}$  with the first line of Eq. \eqref{eqmo},
using $[D_{\mu},D_{\nu}]=g(\partial_{\mu} A_{\nu}-\partial_{\nu} A_{\mu})=gF_{\mu\nu}$, and substituting the $\lambda$ equation of motion, gives the secondary constraint
\begin{equation}\label{secconst}
-g\gamma^{\mu\nu\rho}F_{\mu\nu}\psi_{\rho}=2m^2 \gamma^{\nu}\psi_{\nu}~~~.
\end{equation}
We see that  in the  limit $gF_{\mu\nu}/m^2\to 0$, this
constraint reduces to the simplified form $0= \gamma^{\nu}\psi_{\nu}$, which is familiar
as a gauge constraint in the free Rarita-Schwinger theory.  The four-component covariant equations of this
section will be useful when we turn to path integral quantization and Feynman rules, but
for further analysis it will be simpler to work with the left-chiral reduction of the
covariant equations, to which we proceed next.

\section{Left chiral reduction}
Since the left-chiral and right-chiral components are uncoupled, for many calculations it is simpler to work with the
left-chiral reduction of the model.  This is obtained by converting the action to two-component form for the left-chiral components of $\psi^{\mu}$ and $\lambda$, using the Dirac matrices and left-chiral projector $P_L$ given in  Appendix A of \cite{adlerRS1}.   Defining the two-component four-vector spinor $\Psi^{\mu}$ and its adjoint $\Psi_{\mu }^{\dagger}$, and the two-component spinor
$\ell$ and its adjoint $\ell^{\dagger}$, by
\begin{align}\label{Psidef}
 P_L \psi_{\mu}=&\left( \begin{array} {c}
 \Psi_{\mu}  \\
 0 \\  \end{array}\right) ~~~,\cr
 \psi_{\mu}^{\dagger} P_L=&\Big(\Psi_{\mu }^{\dagger}~~~0\Big)~~~,\cr
  P_L \lambda=&\left( \begin{array} {c}
\ell  \\
 0 \\  \end{array}\right) ~~~,\cr
 \lambda^{\dagger} P_L=&\Big(\ell^{\dagger}~~~0\Big)~~~,\cr
\end{align}
the  action is reduced to its left-chiral part, with Dirac gamma matrices replaced by Pauli spin matrices.
We find
\begin{align}\label{twocomp}
S=&S(\Psi_{\mu})+S(\ell)+S_{\rm interaction}~~~,\cr
S(\Psi_{\mu})=&\int d^4x(-\Psi_0^{\dagger} \vec \sigma \cdot \vec D \times \vec \Psi+ \vec{\Psi}^{\dagger}\cdot  \vec \sigma \times \vec D \Psi_0
+ \vec{\Psi}^{\dagger}\cdot \vec D \times \vec \Psi- \vec{\Psi}^{\dagger} \cdot \vec \sigma \times D_0 \vec \Psi)~~~,\cr
S(\ell)=&i\int d^4x \ell^{\dagger} (D_0-\vec \sigma \cdot  \vec D) \ell~~~,\cr
S_{\rm interaction}=&im\int d^4x(-\ell^{\dagger} \Psi_0 +\ell^{\dagger} \vec \sigma \cdot \vec \Psi + \Psi_0^{\dagger} \ell - \vec{\Psi}^{\dagger} \cdot \vec \sigma \ell)~~~.\cr
\end{align}
Varying $S$ with respect to $\vec{\Psi}^{\dagger}$  and  $\ell^{\dagger}$ gives the Euler-Lagrange
equations
\begin{align}\label{euler}
0=\vec V\equiv &\vec \sigma \times \vec D \Psi_0+ \vec D \times \vec{\Psi}-\vec \sigma \times D_0 \vec \Psi - im \vec \sigma \ell~~~,\cr
0=& (D_0-\vec \sigma  \cdot \vec D) \ell+ m (\vec \sigma \cdot \vec{\Psi} - \Psi_0)~~~,\cr
\end{align}
while varying $S$ with respect to $\Psi_0^{\dagger}$ gives the primary constraint in the form
\begin{equation}\label{primary1}
0=\chi\equiv \vec \sigma \cdot \vec D \times \vec \Psi - im \ell~~~.
\end{equation}

Contracting the first line of Eq. \eqref{euler} with $g^{-1}\vec D$,  using $\vec D \times \vec D=-ig \vec B$, $[\vec D,D_0]=-ig\vec E$, and  $D_0 \chi=0$,
 gives the secondary constraint in the form
\begin{equation}\label{secondary1}
0=\vec \sigma \cdot \vec B \Psi_0-(\vec B + \vec \sigma \times \vec E) \cdot \vec \Psi + g^{-1}m(D_0 -\vec \sigma \cdot \vec D) \ell~~~.
\end{equation}
Substituting the second line of Eq. \eqref{euler} (the Euler-Lagrange equation for $\ell$) into Eq. \eqref{secondary1} turns the secondary constraint into an equation relating $\Psi_0$ to
$\vec \Psi$,
\begin{equation}\label{eliminate}
\Psi_0=\Psi_0[\vec \Psi] = (m^2+g \vec \sigma \cdot \vec B)^{-1}[g(\vec B + \vec \sigma \times \vec E) \cdot \vec \Psi + m^2 \vec \sigma \cdot \vec \Psi]~~~.
\end{equation}
In the zero field limit, this reduces to $\Psi_0= \vec \sigma \cdot \vec \Psi$, which is the left-chiral projection of the four-component relation
$\gamma^{\nu}\psi_{\nu}=0$.

A simplified form of the equation of motion for $\vec \Psi$ is obtained by forming $\vec \sigma \times \vec V  - i\vec V$, which gives
\begin{equation}\label{firstpart}
D_0 \vec \Psi=\vec D \Psi_0+ \frac{1}{2}[-\vec \sigma \times (\vec D \times \vec \Psi) + i \vec D \times \vec \Psi -m \vec \sigma \ell\ ]~~~,
\end{equation}
and substituting  $\vec \sigma \chi$, which becomes
\begin{equation}\label{secondpart}
-\vec \sigma \times (\vec D \times \vec \Psi)=i\vec D \times \vec \Psi + m \vec \sigma \ell~~~,
\end{equation}
leading  to
\begin{equation}\label{simplified}
D_0\vec \Psi=\vec D \Psi_0 + i \vec D \times \vec \Psi~~~.
\end{equation}
As  check, we can verify that Eq. \eqref{simplified} together with  the primary constraint of Eq. \eqref{primary1}  imply the $\ell$ field equation of motion, with $\Psi_0$ related to
$\vec \Psi$ by Eq. \eqref{eliminate}.  Form $\vec D \times D_0 \vec \Psi$ using Eq. \eqref{simplified},  and then form $im D_0 \ell$ using the primary constraint;
combining these and some algebra then gives
\begin{equation}\label{check}
0=im[(D_0 -\vec \sigma \cdot \vec D) \ell+ m(\vec \sigma \cdot \vec \Psi-\Psi_0[\vec \Psi])] ~~~.
\end{equation}
The fact that the $\ell$ field equation of motion is a consequence of the $\vec \Psi$ field equation of motion and constraints  will be used subsequently to expedite the analysis.

\section{Canonical momenta, constraint brackets, and counting of degrees of freedom}
The canonical momentum $\vec P$ conjugate to $\vec \Psi$ and the canonical momentum $P$ conjugate to $\ell$ are defined by
\begin{equation}\label{canondef}
\vec P = \frac{\partial^L S}{\partial(\partial_0\vec \Psi)}~~,~~~P=\frac{\partial^L S}{\partial(\partial_0\ell)}~~~.
\end{equation}
From the action of Eq. \eqref{twocomp} written as
\begin{equation}\label{twocomp1}
S=\int d^4x (-\vec{\Psi}^{\dagger}\times \vec \sigma \cdot  \partial_0 \vec \Psi +i \ell^{\dagger} \partial_0 \ell + {\rm terms~with~no~time~derivatives})
~~~,
\end{equation}
we see that the canonical momenta are given by
\begin{equation}\label{momemta}
\vec P= \vec{\Psi}^{\dagger} \times \vec \sigma~~,~~~P=-i \ell^{\dagger}~~~,
\end{equation}
which can be inverted to give
\begin{equation}\label{invert}
\vec{\Psi}^{\dagger}=\frac{1}{2}(i\vec P-\vec P \times \vec \sigma)~~,~~~\ell^{\dagger}=iP~~~.
\end{equation}
From the canonical bracket definitions \cite{henneaux1}
\begin{align}\label{brac4}
[\Psi_{i\alpha}(\vec x),P_{j\beta}(\vec y)] =&-\delta_{ij}\delta_{\alpha\beta} \delta^3(\vec x- \vec y)~~~,\cr
[\ell_{\alpha}(\vec x),P_\beta(\vec y)]=&-\delta_{\alpha\beta} \delta^3(\vec x- \vec y)~~~,
\end{align}
(with $\alpha, \,\beta$ denoting spinor indices, which will be suppressed henceforth),
we find that the primary constraint $\chi = \vec \sigma \cdot \vec D \times \vec \Psi - im \ell  $ and its adjoint $\chi^{\dagger}=  {\vec \Psi}^{\dagger}\times \vec \sigma  \cdot \overleftarrow D + i m \ell^{\dagger}= \vec P \cdot \overleftarrow D  - m P$ obey the bracket\footnote{Since the standard  Dirac theory of constraints assumes a Lagrangian
with no explicit time dependence \cite{henneaux1},  and hence  time-independent constraints, we will take the external fields $\vec E, \,\vec B$ to be time-independent.  This restriction can
be avoided by treating the gauge fields, as well as the fermions, as dynamical quantized fields.}
\begin{equation}\label{constbrack}
[\chi(\vec x),\chi^{\dagger}(\vec y)]=-i\big(m^2+g \vec \sigma \cdot \vec B(\vec x)\big) \delta^3(\vec x-\vec y)~~~.
\end{equation}
This shows that irrespective of whether the external fields are zero or nonzero, the primary constraints $\chi, \, \chi^{\dagger}$
are second class in the Dirac terminology.  As a consequence of this, the degree of freedom counting does not change discontinuously when the external field
is turned on or off, and Eq. \eqref{constbrack} also shows that the constraint bracket is invertible for small enough $g$ when $m\neq 0$, allowing the interacting theory
to be developed in a perturbation expansion.

 We can now apply the standard formula for counting degrees of freedom \cite{henneaux1},
\begin{equation}\label{degfreedom}
{\rm degrees~of~freedom}=\frac{1}{2}(N-2F-S)~~~,
\end{equation}
in which $N$ is the number of real canonical variables, $F$ is the number of real first class constraints, and $S$ is the number
of real second class constraints.  In our case we have $N_{\vec \Psi}=3 \times 2 \times 2=12$,  $N_{\ell}=2\times 2=4$,    $F=0$, and $S=2 \times 2=4$, giving 6 for the number
of  degrees of freedom for the combined left-chiral $\vec \Psi$ and $\ell$ fields.  This counting of degrees of freedom will be  confirmed in the next section.

\section{Zero external field plane wave solutions}

Let us now consider the case when the external fields $\vec B$ and $\vec E$  are zero, so that the equations of motion for $\vec \Psi$ and $\ell$ become linear
partial differential equations.  We look for plane wave solutions of the form
\begin{equation}\label{planewave}
\vec \Psi= \vec C e^{i\Omega t + i K z}~~,~~~\ell=L e^{i\Omega t + i K z}~~~.
\end{equation}
With this ansatz, and using the zero field limit $\Psi_0=\vec \sigma \cdot \vec \Psi$ of Eq. \eqref{eliminate}, the equations of motion and  constraints become
\begin{align}\label{eqmoplane}
\Omega \vec C=&K \hat z \vec \sigma \cdot \vec C+ i K \hat z \times \vec C~~~,\cr
iK \vec \sigma \cdot \hat z \times \vec C=& im L~~~,\cr
(\Omega-\sigma_3 K)L=&0~~~,\cr
\end{align}
with $\hat z$ a unit vector along the $z$ axis.
Since we saw above that the $l$ equation of motion is a consequence of the $\vec \Psi$ equation of motion and constraints, we expect the third equation
in Eq. \eqref{eqmoplane} to be a consequence of the first two, and we can verify this explicitly.  Taking the cross product of $\hat z$ with the first equation,
and then contracting with $i K \vec \sigma$  we get
\begin{equation}\label{crossfirst}
\Omega iK  \vec \sigma \cdot \hat z \times \vec C= -K^2\vec \sigma \cdot (\hat z \times (\hat z \times \vec C))~~~.
\end{equation}
Substituting the second equation into the left-hand side, this becomes
\begin{equation}\label{crosssecond}
\Omega im L= -K^2\vec \sigma \cdot (\hat z \times (\hat z \times \vec C))~~~.
\end{equation}
But multiplying the second equation by $K \sigma_3 = K \vec \sigma \cdot \hat z$ gives
\begin{equation}\label{altsecond}
imK\sigma_3 L= iK^2 (\vec \sigma \cdot \hat z) (\vec \sigma \cdot \hat z \times \vec C)=-K^2\vec  \sigma \cdot (\hat z \times (\hat z \times \vec C))~~~.
\end{equation}
Subtracting Eq. \eqref{altsecond} from Eq. \eqref{crosssecond} gives
\begin{equation}\label{subract}
im(\Omega-K\sigma_3) L=0~~~,
\end{equation}
which when $m\neq0$ gives the third equation in Eq. \eqref{eqmoplane}.  Hence in looking for zero external field plane wave solutions we need only solve the
first two equations in Eq. \eqref{eqmoplane}.

We now consider two cases: (i) $C_1=C_2=0$ and $C_3 \neq 0$,  and (ii) $C_1 \neq 0$ or $C_2 \neq 0$ with  $C_3$ either zero or nonzero, with the x, y, z axes
numbered respectively 1, 2, 3.  Writing $\vec C_{\perp}=C_1 \hat x + C_2 \hat y$, case (i) corresponds to  $ \vec C_{\perp}=0$ and case (ii) corresponds to
$\vec C_{\perp}\neq 0$.

We begin with case (i), which is easier to analyze.  The second equation of Eq. \eqref{eqmoplane} becomes $L=0$, while the  first equation,
after factoring away $\hat z$,  becomes
\begin{equation}\label{long1}
\Omega C_3 = K \sigma_3 C_3~~~,
\end{equation}
with the solutions
\begin{align}\label{long2}
C_3=&\chi_{\uparrow}~~,~~~ \Omega=K~~~,\cr
C_3=&\chi_{\downarrow}~~,~~~ \Omega=-K~~~,\cr
\end{align}
where we have introduced the notation for Pauli spinors
\begin{align}\label{pauli}
\chi_{\uparrow}=&\left( \begin{array} {c} 1 \\ 0 \\  \end{array}\right)~~~,\cr
\chi_{\downarrow}=&\left( \begin{array} {c} 0 \\ 1 \\  \end{array}\right)~~~.\cr
\end{align}

Next we examine case (ii), for which the transverse part of  the first equation of Eq. \eqref{eqmoplane} becomes
\begin{equation}\label{trans1}
\Omega (\hat x C_1 + \hat y C_2)=iK(\hat y C_1-\hat x C_2)~~~.
\end{equation}
with the solutions
\begin{align}\label{trans2}
C_2=&iC_1,~~~\Omega=K~~~,\cr
C_2=&-iC_1,~~~\Omega=-K~~~.\cr
\end{align}
In terms of $C_1$ and $C_2$,   the second equation of Eq. \eqref{eqmoplane} becomes
\begin{equation}\label{long3}
L=\frac{K}{m} (\sigma_2 C_1-\sigma_1 C_2) ~~~.
\end{equation}
When  $C_1=\chi_{\uparrow}$ for $\Omega=K$, or
$C_1=\chi_{\downarrow}$ for $\Omega=-K$, Eq. \eqref{long3} gives $L=0$.  But when
$C_1=\chi_{\downarrow}$ for $\Omega=K$, Eq. \eqref{long3} gives
\begin{equation}\label{long4}
L=-i\frac{K}{m}\chi_{\uparrow}~~~,
\end{equation}
and when  $C_1=\chi_{\uparrow}$ for $\Omega=-K$, Eq. \eqref{long3} gives
\begin{equation}\label{long5}
L=i\frac{K}{m}\chi_{\downarrow}~~~,
\end{equation}
These last two solutions are not true eigenvectors, but rather Jordan canonical form eigenvectors, as
detailed in Appendix B.
The 6 plane wave solutions that we have found are summarized in Table 2.

\begin{table} [ht]
\caption{The 6 plane wave modes.}
\centering
\begin{tabular}{ c c c c c c c  }
\hline\hline
{\rm eigenvector} & $~~C_1$ & $~~C_2$ & $~~C_3$ & $L$ & $[W]\times {\rm eigenvector}$ &$ ~~~{\rm eigenvalue}=\frac{\Omega}{K}$ \\
$v_1$ &0 & 0 & $\chi_{\uparrow}$ & 0& $v_1$ & 1 \\
$v_2$ &0 & 0 & $\chi_{\downarrow}$ & 0& $-v_2$& $-1$ \\
$v_3$ &$\chi_{\uparrow}$ & $i\chi_{\uparrow}$ & 0 & 0& $v_3$&1 \\
$v_4$ &$\chi_{\downarrow}$ & $-i\chi_{\downarrow}$& 0 & 0&$-v_4$ & $-1$ \\
$v_5$ &$\frac{1}{2}\chi_{\downarrow}$ &$\frac{1}{2} i\chi_{\downarrow}$& 0 & $-i(K/m)\chi_{\uparrow}$  &$v_5+v_1$&1\\
$v_6$ &$\frac{1}{2}\chi_{\uparrow}$&$ -\frac{1}{2}i\chi_{\uparrow}$& 0& $i(K/m)\chi_{\downarrow}$ &$-v_6+v_2$& $-1$\\
\hline
\hline\hline
\end{tabular}
\label{modes}
\end{table}

\section{Wave velocity for the coupled model in an external gauge field}

We turn next to an analysis of wave propagation in an external gauge field, to determine whether there are superluminal modes.  As in \cite{adlerRS1} we employ the
method of characteristics, in which one studies the propagation of wavefronts, that is discontinuities in $\ell$ and the first derivatives of $\vec \Psi$, in the neighborhood of a fiducial point around which  $\vec \Psi$  and  the external fields  are continuous.  Dropping terms that do not contribute to the equations governing  discontinuities,
the $\vec \Psi$ field equation of motion of Eq. \eqref{simplified}  and primary constraint of Eq. \eqref{primary1} become
\begin{align}\label{psifieldeq}
\partial_0\vec \Psi=&\vec \nabla \vec R \cdot \vec \Psi+ i \vec \nabla \times \vec \Psi~~~,\cr
\vec \sigma \cdot \vec \nabla \times \vec \Psi=&im\ell~~~.\cr
\end{align}
We do not have to separately consider the $\ell$ equation of motion because we have seen that this is a consequence of Eqs. \eqref{psifieldeq}.

Absorbing the gauge field coupling $g$ into the definitions of $\vec B$ and $\vec E$, the quantity $\vec R$ is given by
\begin{align}\label{rdef}
\vec R=&(m^4-\vec B^2)^{-1}[m^4 \vec \sigma-i m^2 \vec \sigma \times (\vec B+i\vec E)-\vec Q]~~~,\cr
\vec Q=&\vec B \times \vec E +\vec B \vec \sigma \cdot (\vec B + i \vec E)- i \vec B \cdot \vec E \vec \sigma~~~.\cr
\end{align}
Since we are treating $\vec R$ as a constant, Eqs. \eqref{psifieldeq} form a linear system of equations, which
can be studied by making the Fourier ansatz of Eq. \eqref{planewave}, giving
\begin{equation}\label{psifoureq}
0=\vec F \equiv (\Omega \vec C - i K \hat z \times \vec C)-(m^4-\vec B^2)^{-1}K \hat z [m^4 \vec \sigma-i m^2 \vec \sigma \times (\vec B+i\vec E)-\vec Q]\cdot \vec C~~~,
\end{equation}
with $L$ determined in terms of $C_{1,2}$ by
\begin{equation}\label{ldeterm}
L=\frac{K}{m}(\sigma_1 C_2-\sigma_2 C_1)~~~.
\end{equation}

To solve Eq. \eqref{psifoureq} for $\vec C$, we split it into a part that determines $\vec C_{\perp} \equiv (C_1,\,C_2)$ and a part that determines $C_3$.  These equations become
\begin{align}\label{psifoureq1}
0=&F_3=\Omega C_3  -(m^4-\vec B^2)^{-1} K [m^4 \vec \sigma-i m^2 \vec \sigma \times (\vec B+i\vec E)-\vec Q]\cdot \vec C~~~,\cr
0=&\vec F_{\perp}=(\Omega \vec C_{\perp} - i K \hat z \times \vec C_{\perp})~~~.\cr
\end{align}
The second line of Eq. \eqref{psifoureq1}  implies that
\begin{equation}\label{secondline}
0=(\Omega \vec C_{\perp} - i K \hat z \times \vec C_{\perp})^2=(\Omega^2-K^2) \vec C_{\perp}^2~~~,
\end{equation}
and so $\Omega = \pm K$ and the propagation of modes with $\vec C_{\perp} \neq 0$ is luminal.  The first
line of Eq. \eqref{psifoureq1} can then be solved for the $C_3$ value accompanying these solutions, some of which are eigenvectors in the
Jordan canonical form sense, and $L$ is then
determined by Eq. \eqref{ldeterm}.

For modes with $\vec C_{\perp}= 0$, the first line of Eq. \eqref{psifoureq1} reduces to the set of homogenous equations
\begin{equation}\label{psifoureq3}
0=F_3=\big[\Omega  -(m^4-\vec B^2)^{-1} K [m^4 \sigma_3-i m^2 \big(\vec \sigma \times (\vec B+i\vec E)\big)_3- Q_3] \big]C_3~~~,
\end{equation}
which for $C_3 \neq 0$ requires that
\begin{equation}\label{detlong}
0=\det [\Omega  -(m^4-\vec B^2)^{-1} K [m^4 \sigma_3-i m^2 \big(\vec \sigma \times (\vec B+i\vec E)\big)_3- Q_3]~~~.
\end{equation}
This condition gives a quadratic equation for $\Omega/K$, with  solutions
\begin{align}\label{quadsol}
\frac{\Omega_{\pm}}{K}=& \frac{X\pm \surd{Y}}{\vec B^2-m^4}~~~,\cr
X=&B_1E_2-B_2E_1~~~,\cr
Y=&X^2+(m^4-\vec B^2)(m^4+E_1^2+E_2^2-B_3^2)\cr
=&X^2+(m^4-\vec B^2) (m^4-\vec B^2+E_{\perp}^2+B_{\perp}^2) ~~~,\cr
\end{align}
where $E_{\perp}^2=E_1^2+E_2^2~,~~B_{\perp}^2=B_1^2+B_2^2$.  From this we find
\begin{equation}\label{omprod}
-\frac{\Omega_+}{K} \frac{\Omega_-}{K}=1+\frac{E_{\perp}^2+B_{\perp}^2}{m^4-\vec B^2}~~~,
\end{equation}
which shows that when $m^4-\vec B^2>0$, at least one of the two longitudinal modes must have
$|\Omega/K|>1$ and be superluminal. Implications of this will be discussed in the final section.   When $m=0$, Eq. \eqref{quadsol} reduces to Eq. (B5) of \cite{adlerRS1},
where it is shown that in the standard massless Rarita-Schwinger theory, with no coupling to a spin-$\frac{1}{2}$ field, there are no superluminal modes.

\section{Canonical brackets, Dirac brackets, and Hamiltonian equations of motion}

In Eq. \eqref{momemta} we introduced the canonical momenta $\vec P$ and $P$ conjugate to $\vec \Psi$ and $\ell$, and then computed the constraint bracket $[\chi,\chi^{\dagger}]$.
We can similarly compute the canonical brackets $[\Psi_i, \Psi_j^{\dagger}]$ and $[\ell,\ell^{\dagger}]$ with the results
\begin{align}\label{psibrac}
[\Psi_i(\vec x), \Psi_j^{\dagger}(\vec y)]=&-i\frac{1}{2}\sigma_j\sigma_i\delta^3(\vec x-\vec y)=-i(\delta_{ij}-\frac{1}{2}\sigma_i\sigma_j)\delta^3(\vec x-\vec y)~~~,\cr
[\ell(\vec x),\ell^{\dagger}(\vec y)]=&-i\delta^3(\vec x-\vec y)~~~,\cr
\end{align}
while the canonical brackets of $\Psi_i$ with $\ell^{\dagger}$ and $\ell$ with $\Psi_j^{\dagger}$ vanish.

We turn next to calculating the Dirac brackets of the various quantities.  For any $F(\vec \Psi, \ell)$ and $G(\vec \Psi, \vec{\Psi}^{\dagger}, \ell,\ell^{\dagger})$
the Dirac bracket is given by
\begin{align}\label{diracbrac}
[F(\vec x),G(\vec y)]_D=&[F(\vec x),G(\vec y)]-\int d^3w d^3z[F(\vec x),\chi^{\dagger}(\vec w)]M^{-1}(\vec w,\vec z) [\chi(\vec z),G(\vec y)] ~~~,\cr
M(\vec x,\vec y)=&[\chi(\vec x),\chi^{\dagger}(\vec y)]= -i\big(m^2 + g \vec \sigma \cdot \vec B(\vec x)\big) \delta^3(\vec x-\vec y)~~~.\cr
\end{align}
From this we calculate the following Dirac brackets,
\begin{align}\label{diracbrac1}
[\Psi_i(\vec x),\Psi_j^{\dagger}(\vec y)]_D=& -i\left[(\delta_{ij}-\frac{1}{2}\sigma_i\sigma_j)\delta^3(\vec x-\vec y)- \overrightarrow{D}_{\vec x \, i} \frac{\delta^3(\vec x-\vec y)}{m^2+ g \vec \sigma \cdot \vec B(\vec x)} \overleftarrow{D}_{\vec y \, j}\right]~~~,\cr
[\ell(\vec x),\ell^{\dagger}(\vec y)]_D=&-i\delta^3(\vec x-\vec y)\left[ 1-\frac{m^2}{m^2 + g \vec \sigma \cdot \vec B(\vec x)}\right] =
-i\delta^3(\vec x-\vec y)\frac{ g \vec \sigma \cdot \vec B(\vec x)}{m^2 + g \vec \sigma \cdot \vec B(\vec x)}~~~,\cr
[\ell(\vec x),\Psi_j^{\dagger}(\vec y)]_D=&im\frac{\delta^3(\vec x-\vec y)}{m^2++ g \vec \sigma \cdot \vec B(\vec x)} \overleftarrow{D}_{\vec y \, j}~~~.\cr
\end{align}
The vanishing of the Dirac bracket $[\ell(\vec x),\ell^{\dagger}(\vec y)]_D$ in the $m\to \infty$ limit is a result of the fact that the primary constraint forces
the vanishing of $\lambda$ in this limit, and is also reflected in the form of the $\psi$ field propagator
derived in Sec. 11.  The question of positivity
of the quantum correspondent of these Dirac brackets (obtained by multiplying by $i$ and replacing Dirac brackets with anti-commutators) will be discussed in Sec. 14.

Writing the total action $S$ of Eq. \eqref{twocomp} in the form
\begin{equation}\label{hamform}
S=\int d^4 x (-\vec P \cdot \partial_0 \vec \Psi -P \partial_0 \ell -\Psi_0^{\dagger} \chi - \chi^{\dagger} \Psi_0) - \int dt H~~~,
\end{equation}
we identify the Hamiltonian $H$ as
\begin{equation}\label{hamiltonian}
H= \int d^3x [-\vec{\Psi}^{\dagger} \cdot \vec D \times \vec \Psi+ i \ell^{\dagger} \vec \sigma \cdot \vec D \ell + \vec{\Psi}^{\dagger} \cdot \vec \sigma\times g A_0 \vec \Psi -i \ell^{\dagger} g A_0 \ell
+im(\vec{\Psi}^{\dagger} \cdot \vec \sigma \ell - \ell^{\dagger} \vec \sigma \cdot \vec \Psi)]~~~.
\end{equation}
By construction of the Dirac bracket $[\chi(\vec x), H]_D=0$. So the constraint $\chi(\vec x)$ is a constant, which once zero at an initial time is zero for all
later times.
From Eq. \eqref{hamiltonian} we find, after considerable algebra,
\begin{align}\label{hameqmo}
[\Psi_i(\vec x),H]_D=&-\frac{1}{2}i \sigma_i \chi(\vec x) -gA_0(\vec x) \Psi_i(\vec x) + i\big(\vec D_{\vec x} \times \vec \Psi(\vec x)\big)_i\cr+&
D_{\vec x \, i}\Psi_0[\vec \Psi(\vec x)]~~~,\cr
[\ell(\vec x),H]_D=&-g A_0(\vec x) \ell(\vec x) +\vec \sigma \cdot \vec D_{\vec x} \ell(\vec x) +m\big(\Psi_0[\vec \Psi(\vec x)] - \vec \sigma \cdot \vec \Psi(\vec x)\big)~~~,\cr
\end{align}
where
\begin{equation}\label{psi0psi}
\Psi_0[\vec \Psi(\vec x)]= \int d^3y \frac{\delta^3(\vec x-\vec y)}{m^2 + g \vec \sigma \cdot \vec B(\vec x)}[m^2 \vec \sigma + g \big(\vec B(\vec y) + \vec \sigma \times \vec E(\vec y)\big)] \cdot \vec \Psi(\vec y)~~~
\end{equation}
is Eq. \eqref{eliminate}   with the right-hand side written out in full.
Thus comparing with Eqs. \eqref{euler}   and   \eqref{simplified}, we see that  modulo a term
proportional to the vanishing constraint $\chi$, the Dirac brackets of $\vec \Psi$ and $\ell$ with $H$ reproduce their time development equations of motion.

\section{Path integral quantization}

Path integral quantization of theories with time-independent second class constraints has been addressed by Senjanovic \cite {senj} and Fradkin and
Fradkina \cite{frad}.   Continuing for the moment to work in the left chiral sector, their recipe gives
\begin{align}\label{pathint1}
<{\rm out}|S|{\rm in}>\propto &\int d\mu e^{i[\int d^4 x (-\vec P \cdot \partial_0 \vec \Psi -P \partial_0 \ell) - \int dt H]}~~~,\cr
d\mu=& \delta(\chi)\delta(\chi^{\dagger}) \left[ \det \left( \begin{array} {c c}
   0~~~ & [\chi,\chi^{\dagger}]   \\
   {[\chi^{\dagger},{\chi}]}  & 0\\
\end{array} \right)\right]^{-1/2}
d\vec \Psi d\vec{\Psi}^{\dagger}d\ell d\ell^{\dagger} ~~~.\cr
\end{align}
For ${\cal A}$ a general $2 \times 2$ matrix ${\cal A}=A_0+\vec {\cal A} \cdot \vec \sigma$,
explicit evaluation shows that
\begin{equation}\label{dysonmoore}
 \det \left( \begin{array} {c c}
  0 & {\cal A} \\
  {\cal A} & 0\\
\end{array} \right)=(\det{\cal A})^2~~~.
\end{equation}
Since $[\chi,\chi^{\dagger}]=[\chi^{\dagger},{\chi}]=-iM$, with $M=(m^2+g \vec \sigma \cdot \vec B)\delta^3(\vec x-\vec y)$,
Eq. \eqref{dysonmoore} shows that
\begin{equation}\label{dysonmoore1}
\left[ \det \left( \begin{array} {c c}
   0~~~ & [\chi,\chi^{\dagger}]  \\
   {[\chi^{\dagger},{\chi}]}  & 0~~~\\
\end{array} \right)\right]^{-1/2}= (\det M)^{-1}~~~.
\end{equation}
A further simplification of Eq. \eqref{pathint1} is obtained by noting that
\begin{equation}\label{deltas}
 \delta(\chi)\delta(\chi^{\dagger}) \propto \int d\Psi_0 d\Psi_0^{\dagger} e^{i[\int d^4 x (-\Psi_0^{\dagger} \chi - \chi^{\dagger} \Psi_0)]}~~~,
\end{equation}
allowing us to write
\begin{align}\label{pathint2}
<{\rm out}|S|{\rm in}>\propto &\int d\mu e^{iS}~~~,\cr
d\mu=&(\det M)^{-1}d\Psi_0 d\Psi_0^{\dagger}
d\vec \Psi d\vec{\Psi}^{\dagger}d\ell d\ell^{\dagger} ~~~,\cr
\end{align}
with $S$ in the exponent the total action of Eq. \eqref{twocomp}.
Going back to the full four component action, the path integral takes the form
\begin{align}\label{pathint3}
<{\rm out}|S|{\rm in}>\propto &\int d\mu e^{iS}~~~,\cr
d\mu=&(\det M)^{-1}d\Psi_{\mu} d\Psi_{\mu}^{\dagger}
d\lambda d\lambda^{\dagger} ~~~,\cr
\end{align}
with
$S$ in the exponent the action as written in Eq. \eqref{covariant},
and with $\vec \sigma$ in $M$ replaced by the Dirac matrix  ${\rm diag} (\vec \sigma,\vec \sigma)=\vec \sigma\, 1$.
Equation \eqref{pathint3} will be the starting point for a calculation of the chiral anomaly in
the model with the Rarita-Schwinger field coupled to a spin-$\frac{1}{2}$ field.

\section{The ghost contribution to the chiral anomaly}

Since we have seen in Sec. 5 that the interacting theory can be developed in a perturbation expansion in the gauge coupling $g$, we expect
there to be a well-defined chiral anomaly.  Calculation of this anomaly is the principal aim of the remainder of this paper.  The anomaly
is the sum of two distinct contributions, one from triangle diagrams involving propagation of the fermion fields $\psi_{\mu}$ and $\lambda$, and one from triangle diagrams
involving propagation of bosonic spin-$\frac{1}{2}$ ghosts that enforce the constraints.  Since the later are algebraically much simpler than the former, we will discuss the
ghost contribution to the anomaly first, before turning to the more complex calculations of the spin-$\frac{3}{2}$ contribution in subsequent sections.

The first thing to note about ghost triangles is that there are two sources of multiplicative factors $-1$ in their anomalies, relative to the standard
spin-$\frac{1}{2}$ chiral anomaly.  The first comes from the fact that ghosts are bosons, and so lack the closed loop factor of $-1$ present for fermion
closed loops.  The second is that the ghosts can either have the same chiral transformation properties as the physical fermion fields, or the opposite
chiral transformation properties, and in the latter case there is an additional factor of $-1$ in their anomaly contribution.

With this this in mind, let us recall the heuristic explanation given by Alvarez-Gaum\'e and Witten \cite{witten} of the ghost counting rules for spin-$\frac{3}{2}$
given by Nielsen \cite{nielsen}.  To quote from \cite{witten}:  ``In the quantization of the Rarita-Schwinger field, it is necessary to introduce several
spin-$\frac{1}{2}$ Faddeev-Popov ghosts.  Specifically, one needs two ghosts of the same chirality as $\psi_{\mu}$ and one of opposite chirality. Although
this counting of ghosts sounds odd at first, it really has a simple explanation.  Consider a physical propagating spin-$\frac{3}{2}$ particle of
momentum $k_{\mu}$. The constraints $k_{\mu}\psi^{\mu}=0$ and the gauge invariance under $\psi^{\mu} \to \psi^{\mu}+ k^{\mu} \alpha$ (for any  spin-$\frac{1}{2}$ field
$\alpha$) remove two spin-$\frac{1}{2}$ degrees of freedom of the same chirality as $\psi_{\mu}$.  The additional constraint $\gamma^{\mu} \psi_{\mu}=0$ removes one
spin-$\frac{1}{2}$ degree of freedom of {\it opposite} chirality to $\psi_{\mu}$.  These conditions leave only a physical spin-$\frac{3}{2}$ particle.''

In terms of anomaly contributions, the above reasoning implies a contribution of $-2$ times the usual spin-$\frac{1}{2}$ chiral anomaly from the two same chirality
ghosts, and a contribution of $+1$ times the usual anomaly from the opposite chirality ghost, giving a total anomaly contribution of $-1$.   Now contrast this with what happens in the coupled model of this paper.
Since there are neither a transversality condition nor a fermionic gauge invariance, there is no analog of the two same chirality ghosts and their anomaly contribution
of $-2$.  There is a constraint that is imposed, but this takes the form of Eq. \eqref{spinorform} and involves the removal of a degree of freedom of the {\it same}
chirality as $\psi_{\mu}$ and $\lambda$, rather than of opposite chirality as in the case of the gauge constraint $\gamma^{\mu} \psi_{\mu}=0$.  So the ghost corresponding
to the constraint  of Eq. \eqref{spinorform} will contribute -1 times the standard spin-$\frac{1}{2}$ chiral anomaly.  This gives the same  total $-1$ as comes
from the ghosts in the free spin-$\frac{3}{2}$ theory case, but arises in a very different manner.

Returning now to the path integral of Eq. \eqref{pathint3}, let us try to implement these remarks through a ghost construction.  Introducing a bosonic complex spin-$\frac{1}{2}$ field $\theta(\vec x)$, one can turn the factor $ (\det M)^{-1}$ into a Gaussian by writing
\begin{equation}\label{gaussian1}
(\det M)^{-1} \propto\int d\theta  d\theta^{\dagger}  e^{i\int d^3 x  \theta^{\dagger}(\vec x) \big(m^2+g\vec \sigma \cdot \vec B(\vec x)\big)
\theta(\vec x)}~~~,
\end{equation}
which can be written in a formally Lorentz-covariant way as
\begin{align}\label{gaussian2}
(\det M)^{-1}\propto &\int d\theta d\bar{\theta} e^{i\int d\Sigma_{\mu} \bar{\theta}(\Sigma))G^{\mu} \theta(\Sigma))}~~~,\cr
G^{\mu}=& im^2 \gamma^{\mu}-\frac{1}{2} \epsilon^{\mu\nu\lambda\sigma} \gamma_{\nu}F_{\lambda \sigma}(\Sigma)~~~,\cr
\end{align}
with $d\Sigma^{\mu}$ the volume element of the spacelike surface with coordinate $\Sigma$.  However, neither of these formulas corresponds to a propagating
ghost field $\theta$, and both  yield a ghost anomaly contribution of $0$.

An alternative result is obtained  by treating $M$ as a limit,
\begin{equation}\label{limit}
M \propto  {\rm lim}_{\delta \to 0}\big(m^2+g\vec \sigma \cdot \vec B(\vec x)+ \delta m \gamma^{\mu} D_{\mu}\big)\delta^4(x-y)  ~~~,
\end{equation}
which gives for nonzero $\delta$
\begin{equation}\label{gaussian3}
(\det M)^{-1}\propto \int d\theta d\bar{\theta}  e^{i\int d^4x \bar{\theta}(x)\big(  \gamma^{\mu} D_{\mu}
+m/\delta + (g/(m \delta))\vec\sigma \cdot \vec B(\vec x)\big) \theta(x)}~~~,
\end{equation}
in which $\theta$ is now a bosonic, spin-$\frac{1}{2}$ propagating ghost field with effective mass $m /\delta$.   This ghost will have a triangle anomaly, and since the anomaly is a topological quantity independent of its mass, and unaffected by the coupling to the field $\vec B(\vec x)$, the ghost will give
an anomaly contribution of $-1$ times the standard spin-$\frac{1}{2}$ chiral anomaly, in agreement with what is expected  from the role of the
ghost in implementing the same chirality constraint of Eq. \eqref{spinorform}.

Both of the arguments just given for a ghost contribution of $-1$ are clearly heuristic, so this question invites further investigation using the
formal theory of ghosts in constrained theories \cite{batalin1}, \cite{batalin2}, \cite{marcrev}.

\section{Feynman rules for propagators and vertex parts}

We turn next to calculating the $\psi_{\mu}$ and $\lambda$ field contributions to the chiral anomaly.  The first step is to derive the Feynman rules
for propagator and vertex parts.  Introducing Fourier transforms
\begin{align}\label{fourier1}
\psi_{\mu}(x)=&\frac{1}{(2\pi)^4} \int d^4k e^{ik \cdot x} \psi_{\mu}[k]~~~,\cr
\lambda(x)=&\frac{1}{(2\pi)^4} \int d^4k e^{ik \cdot x} \lambda[k]~~~,\cr
\end{align}
the action of Eq. \eqref{covariant} takes the form (with $\slashed{k}=\gamma^{\nu}k_{\nu}$)
\begin{align}\label{momaction}
S=& \frac{1}{(2\pi)^4} \int d^4k e^{ik \cdot x} S[k]~~~,\cr
S[k]=& -i \bar{\psi}_{\mu}[k]\gamma^{\mu\nu\rho} k_{\nu}\psi_{\rho}[k]-i\bar{\lambda}[k]\slashed{k}\lambda[k]
+m\big(\bar{\lambda}[k]\gamma^{\nu}\psi_{\nu}[k]-  \bar{\psi}_{\nu}[k]\gamma^{\nu}\lambda[k]\big)~~~\cr
=&\big( \bar{\psi}_{\mu}[k]~~\bar{\lambda}[k]\big) \cal{M} \left( \begin{array} {c}
 \psi_{\rho}[k]  \\
 \lambda[k] \\  \end{array}\right)~~~,\cr
\end{align}
with ${\cal M}(k)$ the matrix
\begin{equation}\label{matrixM}
{\cal M}= \left( \begin{array} {c c }
-i  \gamma^{\mu\nu\rho} k_{\nu}   & -m \gamma^{\mu}    \\
 m \gamma^{\rho}    &  -i\slashed{ k} \\  \end{array}\right)~~~.
\end{equation}

The propagator for the coupled $\psi_{\mu}$ and $\lambda$ fields is the matrix ${\cal N}(k)$ that is inverse to ${\cal M}(k)$,
\begin{align}\label{Minverse}
{\cal M}{\cal N}=&  \left( \begin{array} {c c }
\delta^{\mu}_{\sigma}   &0    \\
 0  & 1 \\  \end{array}\right)~~~,\cr
 {\cal N}=& \left( \begin{array} {c c }
N_{1\rho\sigma}   & N_{2\rho}    \\
N_{3\sigma}    &N_4\\  \end{array}\right)~~~,\cr
\end{align}
which can be solved to give the unique answer
\begin{align}\label{Nformula}
N_{1\rho\sigma}=& \frac{-i}{2 k^2}\left[\gamma_{\sigma} \slashed{k} \gamma_{\rho}+ 2\left( \frac{1}{m^2}-\frac{2}{k^2}\right)k_{\rho}k_{\sigma}\slashed{k}\right]~~~,\cr
N_{2\rho}=&\frac{\slashed{k}k_{\rho}}{mk^2}~~~,\cr
N_{3\sigma}=&-\frac{\slashed{k}k_{\sigma}}{mk^2}~~~,\cr
N_4=&0~~~.\cr
\end{align}
Two properties of ${\cal N}$ are worth noting.  First, the fact that $N_4=0$  mirrors the vanishing of the zero external field Dirac bracket $[\ell,\ell^{\dagger}]_D$.
Second, when $m \to \infty$, the off diagonal elements $N_{2\rho}$ and $N_{3\sigma}$ vanish, and $N_{1 \rho\sigma}$ simplifies to
\begin{equation}\label{simplified1}
\tilde{N}_{\rho\sigma}(k)= \frac{-i}{2 k^2}\left[\gamma_{\sigma} \slashed{k} \gamma_{\rho}-\frac{4}{k^2}k_{\rho}k_{\sigma}\slashed{k}\right]~~~,
\end{equation}
which obeys
\begin{equation}\label{simplified2}
\gamma^{\rho}\tilde{N}_{\rho\sigma}=\tilde{N}_{\rho\sigma}\gamma^{\sigma}=0~~~.
\end{equation}
This is a reflection of the fact that as $m\to \infty$, the secondary constraint of Eq. \eqref{secconst} reduces to $\gamma^{\rho}\psi_{\rho}=0$.
For comparison with these formulas, we note that  in our conventions (which have been chosen to agree with \cite{freedman}), the propagator for
a free spin-$\frac{1}{2}$ fermion is
\begin{equation}\label{usualprop1}
s(k)=i/\slashed{k}~~~.
\end{equation}

From the action of Eq. \eqref{covariant} we find that for a Hermitian vector gauge field carrying index $\nu$,  the vertex part Feynman rule is
\begin{equation}\label{vectorvertex}
{\cal V^{\nu}}= \left( \begin{array} {c c }
-i g \gamma^{\mu\nu\rho}    & 0    \\
0    &  -ig\gamma^{\nu} \\  \end{array}\right)~~~,
\end{equation}
with the free indices
$\mu$ and $\rho$ understood to act on the corresponding indices in row and column vectors standing
to the left and right of ${\cal V}^{\nu}$  in the
same matter as these indices act in Eqs. \eqref{momaction} and \eqref{matrixM}.
Since we will compute the chiral anomaly by evaluating the ${\cal A}{\cal V}{\cal V}$ triangle, which is linear in the axial-vector vertex, we
omit the factor $-i$  in the axial-vector vertex part, which we take as
\begin{equation}\label{axialvertex}
{\cal A^{\nu}}= \left( \begin{array} {c c }
 \gamma^{\mu\nu\rho} \gamma_5   & 0    \\
0    &  \gamma^{\nu}\gamma_5 \\  \end{array}\right)~~~.
\end{equation}
The corresponding vector vertex and axial-vector vertex for a free spin-$\frac{1}{2}$ fermion  are respectively $-ig\gamma^{\nu}$ and $\gamma^{\nu}\gamma_5$.

It is also of interest to contrast Eq. \eqref{simplified1} with the propagator for a gauge-fixed free Rarita-Schwinger field.
Noting the identity $\gamma^{\mu\nu\rho}=\frac{1}{2}(\gamma^{\mu}\gamma^{\nu}\gamma^{\rho}-\gamma^{\rho}\gamma^{\nu}\gamma^{\mu})$, and adding the usual
gauge fixing proportional to $\zeta \bar{\psi}_{\mu}\gamma^{ \mu} \slashed{\partial}\gamma^{\rho}\psi_{\rho}$ to the covariant Lagrangian density,
the gauge fixed propagator  $\hat N_{\rho\sigma}(k)$ is obtained by solving
\begin{equation}\label{usualprop}
-i\big[(\frac{1}{2}+\zeta) \gamma^{\mu} \slashed{k} \gamma^{\rho}-\frac{1}{2} \gamma^{\rho} \slashed{k} \gamma^{\mu}\big] \hat N_{\rho\sigma}= \delta^{\mu}_{\sigma}~~~,
\end{equation}
giving
\begin{equation}\label{freedprop}
\hat N_{\rho\sigma}(k)=\frac{-i}{2k^2} \big[\gamma_{\sigma}\slashed{k}\gamma_{\rho}-\frac{1}{k^2}\big(4+\frac{2}{\zeta}\big)k_{\rho}k_{\sigma}\slashed{k}\big]~~~.
\end{equation}
When $\zeta=-\frac{1}{2}$, the gauged fixed propagator $\hat N$ takes the usual ``reverse index'' form for the Rarita-Schwinger propagator widely used in the
literature; when $\zeta \to \infty$ the propagator $\hat N$ limits to that of $\tilde N$ given in Eq. \eqref{simplified1}.  The vector and axial-vector vertices
corresponding to the wave operator in Eq. \eqref{usualprop} are
\begin{align}\label{usualvertices}
\hat{\cal V}^{\nu}=&-ig\big[(\frac{1}{2}+\zeta) \gamma^{\mu} \gamma^{\nu}  \gamma^{\rho}-\frac{1}{2} \gamma^{\rho} \gamma^{\nu} \gamma^{\mu}\big] ~~~,\cr
\hat{\cal A}^{\nu}=&\big[(\frac{1}{2}+\zeta) \gamma^{\mu} \gamma^{\nu}  \gamma^{\rho}-\frac{1}{2} \gamma^{\rho} \gamma^{\nu} \gamma^{\mu}\big]\gamma_5 ~~~.\cr
\end{align}

\section{Ward identities}

The usual spin-$\frac{1}{2}$ propagator of Eq. \eqref{usualprop1} obeys the vector and axial-vector Ward identities
\begin{align}\label{ward1}
-i\gamma^{\mu} k_{\mu}=& s^{-1}(p+k) - s^{-1}(p)~~~,\cr
-i\gamma^{\mu}\gamma_5 k_{\mu}=&s^{-1}(p+k)\gamma_5 + \gamma_5 s^{-1}(p)~~~.
\end{align}
From the linearity of the wave operators on the left-hand side of Eqs. \eqref{Minverse} and \eqref{usualprop} we see that ${\cal N}$, $\tilde N$  and $\hat N$ obey similar Ward identities, with indices $\mu$ and $\rho$ implicit on the right-hand side,
\begin{align}\label{ward2}
\left( \begin{array} {c c }
 -i\gamma^{\mu\nu\rho} k_{\nu}  & 0    \\
0    &  -i\gamma^{\nu}k_{\nu} \\  \end{array}\right)=&{\cal N}^{-1}(k+p)-{\cal N}^{-1}(p)~~~,\cr
\left( \begin{array} {c c }
 -i\gamma^{\mu\nu\rho}\gamma_5 k_{\nu}  & 0    \\
0    &  -i\gamma^{\nu}\gamma_5 k_{\nu} \\  \end{array}\right)=&{\cal N}^{-1}(k+p)\gamma_5+\gamma_5{\cal N}^{-1}(p)~~~,\cr
-i\gamma^{\mu\nu\rho} k_{\nu} =& \tilde{N}^{-1}(k+p)-\tilde{N}^{-1}(p)~~~,\cr
-i\gamma^{\mu\nu\rho}\gamma_5 k_{\nu} =& \tilde{N}^{-1}(k+p)\gamma_5+\gamma_5\tilde{N}^{-1}(p)~~~.\cr
-i\big[(\frac{1}{2}+\zeta) \gamma^{\mu} \slashed{k} \gamma^{\rho}-\frac{1}{2} \gamma^{\rho} \slashed{k} \gamma_{\mu}\big] =& \hat{N}^{-1}(k+p)-\hat{N}^{-1}(p)~~~,\cr
-i\big[(\frac{1}{2}+\zeta) \gamma^{\mu} \slashed{k} \gamma^{\rho}-\frac{1}{2} \gamma^{\rho} \slashed{k} \gamma_{\mu}\big]\gamma_5  =& \hat{N}^{-1}(k+p)\gamma_5+\gamma_5\hat{N}^{-1}(p)~~~.\cr
\end{align}
We will use these in the chiral anomaly calculations of  the subsequent sections.

\section{Chiral anomaly calculations}

There are now many methods in the literature for calculating the chiral anomaly.  Because the vertices and propagators of the preceding section involve products of three
gamma matrices, to minimize the Dirac algebra  we use the observation in the original papers  \cite{adleranomaly}, \cite{bellanomaly}  that the anomaly results from a failure of shift invariance inside a linearly divergent integral, and follow closely the formulation of this given in \cite{jackiwlectures}.  We begin by using the shift method to calculate the usual spin-$\frac{1}{2}$
chiral anomaly, and then give two calculations of the anomaly in the model with the spin-$\frac{3}{2}$ field $\psi_{\mu}$ coupled to the spin-$\frac{1}{2}$ field
$\lambda$.

\subsection{The standard spin-$\frac{1}{2}$ chiral anomaly by the shift method}

The two triangle diagrams with vector vertices $-ig\gamma_{\sigma}$ and $-ig\gamma_{\tau}$, with respective incoming momenta $k_1$ and $k_2$, coupled to an axial-vector
vertex $\gamma^{\nu}\gamma_5$ with incoming momentum $-(k_1+k_2)$, correspond to the amplitude
\begin{align}\label{amplitude1}
{\cal T}_{\sigma\tau}^{\nu}=& \int \frac{d^4 r}{(2\pi)^4} (-1) \rm{tr} \left[\frac{i}{\slashed{r}+\slashed{k_1}}(-ig\gamma_{\sigma})\frac{i}{\slashed{r}}(-ig\gamma_{\tau})
\frac{i}{\slashed{r}-\slashed{k_2}}\gamma^{\nu}\gamma_5\right] \cr
+& \int \frac{d^4 r}{(2\pi)^4} (-1) \rm{tr} \left[\frac{i}{\slashed{r}+\slashed{k_2}}(-ig\gamma_{\tau})\frac{i}{\slashed{r}}(-ig\gamma_{\sigma})
\frac{i}{\slashed{r}-\slashed{k_1}}\gamma^{\nu}\gamma_5\right]~~~. \cr
\end{align}
Forming the axial-vector divergence $-(k_1+k_1)_{\nu}{\cal T}_{\sigma\tau}^{\nu}$, and substituting $-(\slashed{k_1}+\slashed{k_2})\gamma_5 = (\slashed{r}-\slashed{k_2})\gamma_5 + \gamma_5
(\slashed{r}+\slashed{k_1})$  into the first line and $-(\slashed{k_1}+\slashed{k_2})\gamma_5 = (\slashed{r}-\slashed{k_1})\gamma_5 + \gamma_5
(\slashed{r}+\slashed{k_2})$ into the second line, one gets a sum of four terms, each of which contains only $k_1$ or $k_2$ but not both, and hence vanishes,
since there are not enough external momentum factors to form the pseudoscalar $\epsilon^{\tau\sigma\mu\nu} k_{1\, \mu} k_{2\, \nu}$.  Hence with the the chosen routing
of momenta in the triangle, the axial-vector divergence vanishes.  Since the sum of the two diagrams is symmetric under interchange of the vector vertices, it suffices
to test the single vector divergence $k_1^{\sigma} {\cal T}_{\sigma\tau}^{\nu}$, by substituting $\slashed{k_1}=(\slashed{r}+\slashed{k_1}) - \slashed{r}$ into the
first line and $\slashed{k_1}= \slashed{r}- (\slashed{r}-\slashed{k_1})$  into the second line.  This gives a sum of four terms, two of which contain only $k_2$, and hence
vanish, leaving the other two terms,
\begin{equation}\label{shift1}
k_1^{\sigma} {\cal T}_{\sigma\tau}^{\nu}=ig^2\int \frac{d^4 r}{(2\pi)^4}{\rm tr} \left[ \frac{1}{(\slashed{r}+\slashed{k_1})} \gamma_{\tau} \frac{1}{{\slashed{r}-\slashed{k_2}}} \gamma^{\nu} \gamma_5 - \frac{1}{(\slashed{r}+\slashed{k_2})} \gamma_{\tau} \frac{1}{{\slashed{r}-\slashed{k_1}}} \gamma^{\nu} \gamma_5\right]~~~.
\end{equation}
If we could make the shift of integration variable $r \to r+k_2-k_1$ in the first term of Eq. \eqref{shift1}, the two terms would cancel, but this shift is not permitted
inside a linearly divergent integral.  Following \cite{jackiwlectures} we proceed as follows.  Taking $k_1-k_2$ to be infinitesimal, we can write
Eq. \eqref{shift1} as
\begin{align}\label{shift2}
k_1^{\sigma} {\cal T}_{\sigma\tau}^{\nu}
\simeq  & ig^2\int \frac{d^4 r}{(2\pi)^4} (k_1-k_2)_{\kappa} \frac{\partial}{\partial r_{\kappa}} {\rm tr} \left[ \frac{1}{(\slashed{r}+\slashed{k_2})} \gamma_{\tau} \frac{1}{{\slashed{r}-\slashed{k_1}}} \gamma^{\nu} \gamma_5\right] ~~~.\cr
\end{align}
Let us now make the usual Wick rotation to a Euclidean integration region for $r$, which introduces an overall factor of $i$, and use Stokes theorem, which
for a Euclidean four-dimensional integration over a volume $V$ bounded by a surface $S$ states that
\begin{equation}\label{stokes1}
\int_{\rm V}  d^4r \frac{\partial}{ \partial r_{\kappa}} f(r) = \int_{S} dS^{\kappa} f(r)~~~.
\end{equation}
Applying Eq. \eqref{stokes1} to Eq. \eqref{shift2}, and rationalizing the Feynman denominators, we have
\begin{align}\label{shift3}
k_1^{\sigma} {\cal T}_{\sigma\tau}^{\nu}
\simeq  &\frac{ -g^2}{(2\pi)^4} (k_1-k_2)_{\kappa}\int_S dS^{\kappa}\frac{{\rm tr} \left[ (\slashed{r}+\slashed{k_2}) \gamma_{\tau} (\slashed{r}-\slashed{k_1}) \gamma^{\nu} \gamma_5\right]}{(r+k_2)^2(r-k_1)^2}    ~~~.\cr
\end{align}
The trace in the numerator can be simplified to ${\rm tr}[\big((\slashed{k_1}+\slashed{k_2})\gamma_{\tau} \slashed{r}-\slashed{k_2} \gamma_{\tau} \slashed{k_1}\big)
\gamma^{\nu}\gamma_5]$.  Taking now the surface $S$ to be a large three-sphere of radius $R$, the denominator $(r+k_2)^2(r-k_1)^2\simeq R^4$ and so can be pulled
outside the integral. Since the volume of the sphere is $2\pi^2 R^3$, and noting that $dS^{\kappa}$ is a vector parallel
to $r^{\kappa}$, we  have
\begin{equation}\label{average}
\int_S dS^{\kappa}  {\rm tr}[\big((\slashed{k_1}+\slashed{k_2})\gamma_{\tau} \slashed{r}-\slashed{k_2} \gamma_{\tau} \slashed{k_1}\big)
\gamma^{\nu}\gamma_5]=2\pi^2 R^4  {\rm tr}[(\slashed{k_1}+\slashed{k_2})\gamma_{\tau} (\gamma^{\kappa}/4) \gamma^{\nu}\gamma_5]~~~.
\end{equation}
Thus the $R$ factors cancel out as the sphere radius approaches $\infty$, and we find for the
vector vertex anomaly
\begin{align}\label{finalanomaly}
k_1^{\sigma} {\cal T}_{\sigma\tau}^{\nu}=&\frac{ -g^2}{(2\pi)^4} (k_1-k_2)_{\kappa}2\pi^2{\rm tr}[(\slashed{k_1}+\slashed{k_2})\gamma_{\tau} (\gamma^{\kappa}/4) \gamma^{\nu}\gamma_5]\cr
= &\frac{g^2}{16 \pi^2} {\rm tr}[\slashed{k_1}\gamma_{\tau} \slashed{k_2} \gamma^{\nu}\gamma_5]~~~.\cr
\end{align}
When vector vertex conservation is enforced by adding a polynomial to the amplitude, Eq. \eqref{finalanomaly} yields the usual answer for the axial-vector anomaly.

\subsection{The coupled model anomaly using the propagator $\tilde N$}

We turn next to the calculation of the anomaly in the coupled model.  Using the propagator ${\cal N}$ of Eqs. \eqref{Minverse} and \eqref{Nformula} and the vertices
of Eqs. \eqref{vectorvertex} and \eqref{axialvertex} in the triangle diagram, one gets a polynomial of second degree in $1/m^2$. Since the anomaly is a topological
quantity independent of $m$, only the term in the triangle of zeroth order in $1/m^2$ can contribute, so we can simplify the anomaly calculation by using
the $m \to \infty$ limit of ${\cal N}$ given in Eq. \eqref{simplified1}.  Thus in place of Eq. \eqref{amplitude1} of the preceding subsection, we consider
\begin{align}\label{newamplitude}
\tilde{{\cal T}}_{\sigma\tau}^{\nu}=& \int \frac{d^4 r}{(2\pi)^4} (-1) \rm{tr} [\tilde{N}(r+k_1) {\cal V}_{\sigma} \tilde{N}(r) {\cal V}_{\tau} \tilde{N}(r-k_2){\cal A}^{\nu} \cr  +&\tilde{N}(r+k_2) {\cal V}_{\tau} \tilde{N}(r) {\cal V}_{\sigma} \tilde{N}(r-k_1){\cal A}^{\nu}] ~~~,\cr
\end{align}
where the vector and axial vertex parts are respectively the upper left diagonal elements of Eqs. \eqref{vectorvertex} and \eqref{axialvertex}. Contracting with $i(k_1+k_2)_{\nu}$ to test the axial divergence, and using the respective  Ward identities (with indices $\mu$ and $\rho$ implicit on the right-hand side)
\begin{align}\label{newward1}
i\gamma^{\mu\nu\rho}\gamma_5 (k_1+k_2)_{\nu} =& \tilde{N}^{-1}(r-k_2)\gamma_5+\gamma_5\tilde{N}^{-1}(r+k_1)~~~,\cr
i\gamma^{\mu\nu\rho}\gamma_5 (k_1+k_2)_{\nu} =& \tilde{N}^{-1}(r-k_1)\gamma_5+\gamma_5\tilde{N}^{-1}(r+k_2)~~~,\cr
\end{align}
in the first and second lines of Eq. \eqref{newamplitude}, we get again a sum of four terms, each of which contains only $k_1$ or $k_2$ and so vanish.  So the
axial-vector divergence vanishes.  Contracting with $k_1^{\sigma}$ to test the vector divergence, and using the respective Ward identities (again with indices $\mu$
and $\rho$ implicit on the right)
\begin{align}\label{newward2}
-i\gamma^{\mu\sigma\rho}k_{1\sigma} =& \tilde{N}^{-1}(r+k_1)-\tilde{N}^{-1}(r)~~~,\cr
-i\gamma^{\mu\sigma\rho}k_{1\sigma} =& \tilde{N}^{-1}(r)-\tilde{N}^{-1}(r-k_1)~~~,\cr
\end{align}
in the first and second lines of Eq. \eqref{newamplitude}, we get a sum of four terms, two of which contain only $k_2$ and vanish, leaving the other two terms
\begin{equation}\label{newshift1}
k_1^{\sigma}\tilde{{\cal T}}_{\sigma\tau}^{\nu}=g\int \frac{d^4 r}{(2\pi)^4}\rm{tr} [\tilde{N}(r+k_1)  {\cal V}_{\tau} \tilde{N}(r-k_2){\cal A}^{\nu}
-\tilde{N}(r+k_2) {\cal V}_{\tau}  \tilde{N}(r-k_1){\cal A}^{\nu}] ~~~,
\end{equation}
We can now use the identity
\begin{equation}\label{gammaident}
\gamma^{\alpha\tau\beta}=\frac{1}{2}[\gamma^{\alpha},\gamma^{\tau}]\gamma^{\beta}-\gamma^{\alpha}\eta^{\tau\beta}+\eta^{\alpha\beta}\gamma^{\tau}
\end{equation}
and the projection property of Eq. \eqref{simplified2} to eliminate the terms in Eq. \eqref{gammaident} in which $\gamma^{\beta}$ stands to the right or
$\gamma^{\alpha}$ stands to the left, allowing us to replace the vertex ${\cal V}_{\tau}$ by $-ig \gamma_{\tau}$ and the vertex ${\cal A}^{\nu}$ by $\gamma^{\nu}\gamma_5$, giving the simplified formula
\begin{equation}\label{newshift2}
k_1^{\sigma}\tilde{{\cal T}}_{\sigma\tau}^{\nu}=-ig^2\int \frac{d^4 r}{(2\pi)^4}\rm{tr} [\tilde{N}(r+k_1) \gamma_{\tau} \tilde{N}(r-k_2)\gamma^{\nu}\gamma_5
-\tilde{N}(r+k_2) \gamma_{\tau}  \tilde{N}(r-k_1)\gamma^{\nu}\gamma_5] ~~~.
\end{equation}
Taking $k_1-k_2$ to be infinitesimal, making the Wick rotation to a Euclidean integration region for $r$, and applying Stokes theorem, this gives
\begin{equation}\label{newshift3}
k_1^{\sigma}\tilde{{\cal T}}_{\sigma\tau}^{\nu}\simeq \frac{g^2}{(2\pi)^4}(k_1-k_2)_{\kappa} \int_S dS^{\kappa} \rm{tr} [ \tilde{N}(r+k_2) \gamma_{\tau}  \tilde{N}(r-k_1)
\gamma^{\nu}\gamma_5] ~~~.
\end{equation}
Substituting Eq. \eqref{simplified1} for the $\tilde N$ factors, this becomes
\begin{equation}\label{newshift4}
k_1^{\sigma}\tilde{{\cal T}}_{\sigma\tau}^{\nu}\simeq -\frac{g^2}{(2\pi)^4}(k_1-k_2)_{\kappa} \int_S dS^{\kappa}\frac{T}{(r+k_2)^2 (r-k_1)^2} ~~~,
\end{equation}
with $T$ given by
\begin{align}\label{Tdef}
T=&\frac{1}{4}\rm{tr}\left[ \left(\gamma_{\beta}(\slashed{r}+\slashed{k_2}) \gamma^{\alpha}-\frac{4}{(r+k_2)^2} (r+k_2)_{\beta}(r+k_2)^{\alpha} (\slashed{r} +\slashed{k_2})\right)\gamma_{\tau} \right.\cr \times&\left.\left(\gamma_{\alpha}(\slashed{r}-\slashed{k_1}) \gamma^{\beta}-\frac{4}{(r-k_1)^2} (r-k_1)_{\alpha}(r-k_1)^{\beta} (\slashed{r} -\slashed{k_1})\right)\gamma^{\nu}\gamma_5\right]~~~\cr
=&\left[1+4 \frac{[(r+k_2)\cdot (r-k_1)]^2}{(r+k_2)^2 (r-k_1)^2}\right]{\rm tr} \left[ (\slashed{r}+\slashed{k_2}) \gamma_{\tau} (\slashed{r}-\slashed{k_1}) \gamma^{\nu} \gamma_5\right]\cr
=& [5+O(1/r^2)]{\rm tr} \left[ (\slashed{r}+\slashed{k_2}) \gamma_{\tau} (\slashed{r}-\slashed{k_1}) \gamma^{\nu} \gamma_5\right]~~~.\cr
\end{align}
We obtained this result by calculation ``by hand'', and checked it using the program FeynCalc \cite{feyn}.
Without carrying the computation further, by comparing Eqs. \eqref{newshift4} and \eqref{Tdef} with Eq. \eqref{shift3} of the preceding subsection, we see that the answer for the coupled system anomaly,
excluding the ghost contribution, is 5 times the standard spin-$\frac{1}{2}$ chiral anomaly.  This is the same as the non-ghost part of the model in the uncoupled limit
$m=0$, where the spin-$\frac{3}{2}$ triangle contributes 4 times the standard chiral anomaly \cite{nielsen1} , and the spin-$\frac{1}{2}$ triangle contributes 1 times
the standard chiral anomaly, for a total of 5 times the standard anomaly.

\subsection{An alternative calculation:  the coupled model anomaly using the propagator $\hat N$}

An alternative calculation starts from the triangle for the gauged fixed free massless spin-$\frac{3}{2}$ theory, using the propagator and vertices of Eqs. \eqref{freedprop}
and \eqref{usualvertices}.  Starting from the definition
\begin{align}\label{newnewamplitude}
\hat{{\cal T}}_{\sigma\tau}^{\nu}=& \int \frac{d^4 r}{(2\pi)^4} (-1) \rm{tr} [\hat{N}(r+k_1) \hat{\cal V}_{\sigma} \hat{N}(r) \hat{\cal V}_{\tau} \hat{N}(r-k_2)\hat{\cal A}^{\nu} \cr  +&\hat{N}(r+k_2) \hat{\cal V}_{\tau} \hat{N}(r) \hat{\cal V}_{\sigma} \hat{N}(r-k_1)\hat{\cal A}^{\nu}] ~~~,\cr
\end{align}
and using the Ward identities of Eq. \eqref{ward2}, we find by the same reasoning as used above that the axial divergence is zero, and the vector divergence is given after Wick rotation, Taylor expansion, and application of Stokes theorem, by
\begin{equation}\label{newnewshift1}
k_1^{\sigma}\hat{{\cal T}}_{\sigma\tau}^{\nu}\simeq \frac{ig}{(2\pi)^4}(k_1-k_2)_{\kappa}\int_S dS^{\kappa} \rm{tr} [\hat{N}(r+k_2) \hat{\cal V}_{\tau}  \hat{N}(r-k_1)\hat{\cal A}^{\nu}] ~~~.
\end{equation}
This equation contains many more gamma matrices than appeared in the previous calculations, so we used the program FeynCalc to evaluate it.  The result is that
the anomaly is independent of the gauge fixing parameter $\zeta$, and is 4 times the standard spin-$\frac{1}{2}$ anomaly, in agreement with the result obtained in
\cite{nielsen1} for $\zeta=-1/2$ using the heat kernel regularization method.

To see the relation between this calculation and that of the previous subsection,
let us consider the behavior of the  propagator-vertex product $\hat{N}(r)\hat{\cal V}^{\sigma}$ as $\zeta \to \infty$, since  with the replacements
of $r$ by $r+k_2$ or $r-k_1$ and $\sigma$ by $\tau$ or $\nu$ this gives (up to a factor of $i$ in the axial vertex case) the behavior of the propagator-vertex products appearing in Eq. \eqref{newnewshift1}.
Decomposing the product into a sum of four terms, we write
\begin{equation}\label{prodstudy1}
(\hat{N}(r)\hat{\cal V}^{\sigma})_{\alpha}^{~\delta}=\hat{N}(r)_{\alpha\beta}(\hat{\cal V}^{\sigma})^{\beta\delta}={\rm term ~1}+{\rm term ~ 2}+{\rm term ~3}+{\rm term  ~4}~~~,
\end{equation}
with
\begin{align}\label{fourparts}
{\rm term ~1}=&\frac{-i}{2r^2} \big[\gamma_{\beta}\slashed{r}\gamma_{\alpha}-\frac{4}{r^2}r_{\beta}r_{\alpha}\slashed{r}\big]  (-ig) \gamma^{\beta\sigma\delta}       ~~~,\cr
{\rm term ~ 2}=&\frac{i}{\zeta (r^2)^2}r_{\beta}r_{\alpha}\slashed{r}   (-ig) \gamma^{\beta\sigma\delta}      ~~~,\cr
{\rm term ~3}=&\frac{-i}{2r^2} \big[\gamma_{\beta}\slashed{r}\gamma_{\alpha}-\frac{4}{r^2}r_{\beta}r_{\alpha}\slashed{r}\big]   (-ig)\zeta \gamma^{\beta} \gamma^{\sigma}  \gamma^{\delta}      ~~~,\cr
{\rm term  ~4}=& \frac{i}{\zeta (r^2)^2}r_{\beta}r_{\alpha}\slashed{r} (-ig)\zeta \gamma^{\beta} \gamma^{\sigma}  \gamma^{\delta}  =g\frac{r_{\alpha}}{r^2}  \gamma^{\sigma}\gamma^{\delta}      ~~~. \cr
\end{align}
We see that term 1 equals the propagator-vertex product $\tilde{N}(r)\tilde{\cal V}^{\sigma}$ of the preceding subsection,  and term 3 vanishes by the projection
property of Eq. \eqref{simplified2}.  Also, in the limit $\zeta \to \infty$, term 2 vanishes, and term 4 approaches a constant, which by the projection property annihilates the term 1 coming from the second propagator-vertex product.  Thus Eq. \eqref{newnewshift1} yields the calculation of the preceding section, coming from the product of the two terms 1, plus the product of the two terms 4, giving
\begin{align}\label{subtractionpiece}
k_1^{\sigma}\hat{{\cal T}}_{\sigma\tau}^{\nu}=&{\rm term~1 ~contribution}+
\frac{ig}{(2\pi)^4}(k_1-k_2)_{\kappa}\int_S dS^{\kappa} \frac{\rm{tr} \big[  \big(g(r+k_2)_{\alpha}\gamma_{\tau}\gamma^{\delta} \big) \big( i(r-k_1)_{\delta} \gamma^{\nu} \gamma^{\alpha} \gamma_5  \big)\big]}{(r+k_2)^2(r-k_1)^2} ~~~\cr
=&{\rm term~1~ contribution}+\frac{g^2}{(2\pi)^4}(k_1-k_2)_{\kappa}\int_S dS^{\kappa} \frac{{\rm tr} \left[ (\slashed{r}+\slashed{k_2}) \gamma_{\tau} (\slashed{r}-\slashed{k_1}) \gamma^{\nu} \gamma_5\right]}{(r+k_2)^2(r-k_1)^2}~~~.\cr
\end{align}
Comparing the final term in Eq. \eqref{subtractionpiece}  with Eq. \eqref{shift3}, we see that the term 4 contribution is -1 times the standard spin-$\frac{1}{2}$ anomaly.  Since the total is 4 times the
spin-$\frac{1}{2}$ anomaly, the term 1 contribution is determined to be 5 times the spin-$\frac{1}{2}$ anomaly, confirming the result found by direct
calculation in the preceding subsection.

\section{Discussion and directions for further work}

The calculations of the preceding sections show that the coupled model eliminates some, but not all, of the problems associated with gauged Rarita-Schwinger fields.
Specifically:
 \begin{enumerate}
\item  The coupled model eliminates the discontinuity in the number of degrees of freedom when an external field is present, relative to the zero field case, discussed in detail in \cite{henneaux}.
\item  The non-perturbative behavior of the Dirac bracket, already clear from the zero mass limit of \cite{johnson} and \cite{velo}, and discussed in detail in
\cite{adlerRS1}, \cite{adlerRS2}, and \cite{henneaux} is eliminated.  The coupled model admits a perturbative expansion in powers of the gauge coupling $g$, as well
as permitting calculation of the perturbative chiral anomaly.
\item  The zero external field plane wave solutions in the coupled model are not all eigenvectors of the wave matrix.  Because the wave matrix is non-Hermitian,
as detailed in Appendix B, some of the plane wave solutions are only eigenvectors in the Jordan canonical form sense.
\item  Tachyonic behavior of the longitudinal spin-$\frac{3}{2}$ modes, first noted in \cite{velo}, is not eliminated, unlike the uncoupled Rarita-Schwinger field
studied in \cite{adlerRS1}, where the longitudinal modes are luminal.  This could indicate a problem with the coupled theory, or could indicate that there is an
instability in the presence of gauge fields that triggers dynamical breakdown of chiral symmetry, in which left- and right-chiral modes become coupled through
a conventional mass term.  There is an extensive literature on dynamical chiral symmetry breaking through condensate formation in unified models \cite{peskin}, as well as on dynamical chiral symmetry breaking  in quantum chromodynamics \cite{adlerdavis}.  Because the model studied in this paper contains  both helicity $\pm \frac{3}{2}$ and $\pm \frac{1}{2}$ fields,  chiral symmetry breaking leading to a massive spin-$\frac{3}{2}$ field is kinematically allowed.  An important issue for the
future is to set up the gap equation to study whether dynamical chiral symmetry breaking is realized in the coupled
model.

\item The problem with non-positivity of Dirac brackets (or more precisely, of the corresponding quantum anti-commutators) in the presence of an external field, first noted in \cite{johnson}, verified in \cite{velo}, and discussed in
detail in \cite{henneaux}, is not eliminated.  In the large $m^2$ limit the $\psi$ field brackets are positive semi-definite, but as noted in the early papers
 non-positivity occurs for strong enough external fields.   As first noted  in \cite{velo}, this does not prevent a perturbative expansion in the gauge coupling
  $g$. It also does not appear
to cause a difficulty in formulating path integral quantization.  However, as discussed in \cite{henneaux}, canonical quantization in strong external fields will
require an indefinite metric Hilbert space, and this deserves study.  The brackets for the field $\lambda$ are non-positive for arbitrarily weak fields, and
vanish in the large $m^2$ limit.
\item Although there are internal cross checks on most of the calculations of this paper, including the fermion loop contribution to the chiral anomaly, the
arguments we have given for a ghost contribution of $-1$ as opposed to $0$ are purely heuristic, and need to be supported (or refuted) by better methods.
The restriction to time-independent constraints, which requires a restriction to time-independent external fields $\vec B$ and $\vec E$, may obscure the correct
way to handle the ghost contribution.  This restriction can be eliminated by including a kinetic term for the gauge fields and treating them as dynamical fields,
which will lead to a more complicated system of constraints, with both bosonic and fermionic constraints.
\end{enumerate}

 We believe  the most immediate avenues for further investigation are (i) the study of dynamical chiral symmetry breaking, to see whether the coupled model
  generates a mass term, and  (ii) dealing with the ghost issue, which will determine whether the coupled model chiral anomaly is unchanged from the uncoupled model value,
 as suggested by the calculations of this paper.  Beyond this, it will be important to extend the analysis of this paper to non-Abelian gauge models, such as the model
 of \cite{adlersu8} from which the abelianized model treated in this paper is abstracted.

\section{Acknowledgements}
I wish to thank Marc Henneaux, Pablo Pais, and Edward Witten for helpful emails and/or conversations in the course of my studies of Rarita-Schwinger theory.  The problems found in the collaboration \cite{henneaux}  with Henneaux and Pais  were a direct motivation for pursuing the investigation reported here.  I also wish to thank Pablo Pais for reading through this paper.
The use of FeynCalc \cite{feyn} was suggested by Liang Dai, Matthew Low, Paul Langacker, and Prahar Mitra, and I wish in particular to thank Paul Langacker for assistance in getting it running and sending helpful illustrative examples.  I also wish to acknowledge the hospitality of the Aspen Center for Physics, where this work was started,  and its support by the National Science Foundation under Grant No. PHYS-1066293.

\appendix
\section{Conventions and summary of identities}

We follow the metric and gamma matrix conventions of  \cite{freedman} that were used in \cite{adlerRS1}, \cite{adlerRS2}, \and \cite{henneaux}, but change the
action normalization by a factor of 2 to bring that also into agreement with \cite{freedman}.
We note the following covariant derivative and Pauli matrix identities from \cite{adlerRS1} that we use here:
\begin{align}\label{ident}
\vec D \times \vec D=&\overleftarrow D \times \overleftarrow D= -ig \vec B~~~,\cr
[\vec D,D_0]=&-ig\vec E~~~,\cr
(\vec \sigma \times \vec D)^2=&2 \vec D^2 + g \vec \sigma \cdot \vec B~~~,\cr
(\vec \sigma \cdot \vec D)^2=& \vec D^2 + g \vec \sigma \cdot \vec B~~~,\cr
\vec D \cdot (\vec \sigma \times \vec D)=&ig \vec \sigma \cdot \vec B~~~,\cr
(\vec \sigma \times \overleftarrow D)\cdot \overleftarrow D =&-ig \vec \sigma \cdot \vec B~~~,\cr
\vec  \sigma \times \vec \sigma =&2i\vec \sigma  ~~~,\cr
\vec \sigma \cdot \vec  v\sigma_j=&v_j+i(\vec \sigma \times \vec v)_j~~~,\cr
\sigma_j \vec \sigma \cdot \vec  v=& v_j-i(\vec \sigma \times \vec v)_j~~~,\cr
(\vec \sigma \times \vec v)_i \sigma_j \sigma_i=&2iv_j~~~,\cr
\vec B=i\vec A- \vec A  \times \vec \sigma \leftrightarrow   \vec A=&\frac{1}{2}(\vec B \times \vec \sigma)~~~.
\end{align}
Additionally, we use the following Dirac gamma matrix identities:
\begin{align}\label{ident1}
\gamma^{\mu\nu\rho}=&\frac{1}{2}(\gamma^{\mu}\gamma^{\nu}\gamma^{\rho}-\gamma^{\rho}\gamma^{\nu}\gamma^{\mu})~~~,\cr
=&\frac{1}{2}[\gamma^{\mu},\gamma^{\nu}]\gamma^{\rho}-\gamma^{\mu}\eta^{\nu\rho}+\eta^{\mu\rho}\gamma^{\nu}~~~.\cr
\end{align}

\section{Calculation of plane wave modes}

Writing the first equation in Eq. \eqref{eqmoplane} in the form
\begin{equation}\label{eqmoplanew}
(\Omega/K) [C] = [W] [C]~~~,
\end{equation}
with $[C]$ the colummn vector listing the 6 components of the vector spinor $\vec C$ written
in the order $(C_1^{\uparrow}~~C_1^{\downarrow}~~C_2^{\uparrow}~~C_2^{\downarrow}~~C_3^{\uparrow}~~C_3^{\downarrow})$,
the wave matrix [W] takes the form
\begin{equation}\label{wavematrix}
 [W]= \left[ \begin{array} {c c c c c c}
0  & 0 & -i & 0 & 0 &0  \\
0  & 0 & 0 &-i  & 0 & 0   \\
i  & 0 & 0 & 0 & 0 &  0  \\
0  & i & 0 & 0 & 0 & 0   \\
0  & 1 & 0 & -i & 1 &  0  \\
1  & 0 & i & 0 & 0 & -1   \\
 \end{array}\right] ~~~.
\end{equation}
Since $[W]$ is not self-adjoint, the best one can do is to transform it to Jordan canonical form.
The corresponding right eigenvectors are readily seen to be
\begin{align}\label{righteig}
v_1^T=&( 0 ~~  0 ~~  0 ~~ 0  ~~ 1  ~~ 0  )~~~,\cr
v_2^T=&(0  ~~ 0  ~~ 0  ~~ 0  ~~ 0  ~~ 1  )~~~,\cr
v_3^T=&( 1 ~~ 0  ~~ i  ~~ 0  ~~ 0  ~~ 0  )~~~,\cr
v_4^T=&(0  ~~ 1  ~~ 0  ~~ -i  ~~ 0  ~~ 0  )~~~,\cr
v_5^T=&( 0 ~~ \frac{1}{2}  ~~  0 ~~   \frac{i}{2} ~~ 0  ~~  0 )~~~,\cr
v_6^T=&(  \frac{1}{2} ~~ 0  ~~  -\frac{i}{2}  ~~  0 ~~  0 ~~  0) ~~~,\cr
\end{align}
where we have written the transposes ${}^T$ of the column vectors as row vectors.  Table 2 is constructed
from Eq. \eqref{righteig}.

\end{document}